%% file: main.tex
\RequirePackage[l2tabu,orthodox]{nag}

\documentclass[conference,letterpaper]{IEEEtran}

\usepackage[utf8]{inputenc}
\usepackage[T1]{fontenc}
\usepackage{microtype}
\usepackage[abbreviations]{foreign}
\usepackage{booktabs}
\usepackage[all]{nowidow}
\usepackage{graphicx}
\usepackage{mathtools}
\usepackage{ifthen}
\usepackage{amssymb}
\usepackage{xcolor}
\usepackage{subcaption}
\usepackage[numbers]{natbib}
\usepackage[colorlinks=true,allcolors=blue,breaklinks]{hyperref}
\usepackage{balance}
\usepackage{pgf-umlsd}
\usepackage{url}
\usepackage{tcolorbox}
\usepackage[scaled=.8]{beramono}
\usepackage[english]{babel}
\usepackage{cleveref}
\usetikzlibrary{arrows.meta,automata,positioning}

\newboolean{showcomments}
\setboolean{showcomments}{true}
\ifthenelse{\boolean{showcomments}}
{ \newcommand{\mynote}[3]{
		\fbox{\sffamily\scriptsize#1}
		{\small$\blacktriangleright$\emph{\color{#3}{#2}}$\blacktriangleleft$}}}
{ \newcommand{\mynote}[3]{}}
\def\BibTeX{{\rm B\kern-.05em{\sc i\kern-.025em b}\kern-.08em

	T\kern-.1667em\lower.7ex\hbox{E}\kern-.125emX}}

\begin{document}

\title{MOON: Assisting Students in Completing Educational Notebook Scenarios}

\author{}
\author{
	\IEEEauthorblockN{Christophe Casseau\IEEEauthorrefmark{1}, Jean-Rémy Falleri\IEEEauthorrefmark{1}\IEEEauthorrefmark{2}, Thomas Degueule\IEEEauthorrefmark{1} and Xavier Blanc\IEEEauthorrefmark{1}}
	\IEEEauthorblockA{
		\IEEEauthorrefmark{1}Univ. Bordeaux, Bordeaux INP, CNRS, LaBRI, UMR5800 \\
		Talence, France \\
		name.surname@labri.fr}
	\IEEEauthorblockA{
		\IEEEauthorrefmark{2}Institut Universitaire de France}
}

\IEEEpubid{978-1-6654-4592-4/21/$31.00 ©2023 European Union}

\IEEEoverridecommandlockouts
\IEEEpubid{\makebox[\columnwidth]{978-1-6654-4592-4/21/\$31.00 ©2023 European Union  \hfill} \hspace{\columnsep}\makebox[\columnwidth]{ }}

\maketitle

\IEEEpubidadjcol

\input{abstract}

\input{introduction}
\input{description}
\input{system_design}
\input{implementation}
\input{experimental_design}
\input{user_study}
\input{related_work}
\input{conclusion}

\balance
\bibliographystyle{IEEEtranN}
\bibliography{biblio}

\end{document}

%% file: abstract.tex
\begin{abstract}
% Notebooks are more and more used for teaching purposes
Jupyter notebooks are increasingly being adopted by teachers to deliver interactive practical sessions to their students.
Notebooks come with many attractive features, such as the ability to combine textual explanations, multimedia content, and executable code alongside a flexible execution model which encourages experimentation and exploration.

% Issues with notebooks in a teaching environment
However, this execution model can quickly become an issue when students do not follow the intended execution order of the teacher, leading to errors or misleading results that hinder their learning.
To counter this adverse effect, teachers usually write detailed instructions about how students are expected to use the notebooks.
Yet, the use of digital media is known to decrease reading efficiency and compliance with written instructions, resulting in frequent notebook misuse and students getting lost during practical sessions.

% Our idea to alleviate the raised issues
In this article, we present a novel approach, MOON, designed to remedy this problem.
The central idea is to provide teachers with a language that enables them to formalize the expected usage of their notebooks in the form of a script and to interpret this script to guide students with visual indications in real time while they interact with the notebooks.

% Evaluation & Results
We evaluate our approach using a randomized controlled experiment involving 21 students, which shows that MOON helps students comply better with the intended scenario without hindering their ability to progress.
Our follow-up user study shows that about 75\% of the surveyed students perceived MOON as rather useful or very useful.
\end{abstract}

%% file: introduction.tex
\section{Introduction}

Teachers are increasingly using Jupyter notebooks as a support for educational activities such as graded assignments~\cite{nbgrader, 10.1145/3328778.3366947, su132112050}, interactive textbooks~\cite{10.1145/3408877.3432361}, interactive exercise worksheets, or live coding~\cite{10.1145/3330430.3333627}.
The primary advantage of a notebook lies in its ability to gather all the essential tools within a single platform.
This allows teachers to provide students with an interactive document that includes text and instructions, as well as executable code, images, and videos.
The advantages of using notebooks in an educational context are assessed not only in scientific disciplines such as physics~\cite{sutrini2022potential,suarez2021teaching}, chemistry~\cite{doi:10.1021/acs.jchemed.0c01071, doi:10.1021/acs.jchemed.2c00193, doi:10.1021/acs.jchemed.1c00142, doi:10.1021/acs.jchemed.9b01131, 10.5555/3007225.3007252}, mathematics~\cite{article} or computational sciences~\cite{Vial2018TeachingPT, 8876536, casseau2021immediate, inproceedings}, but also in other disciplines such as humanities for example~\cite{9387480}.

Writing instructions for a notebook-based learning activity is challenging.
It is notoriously difficult to write instructions for digital media as studies have shown that reading efficiency in such media is lower than for paper-based support~\cite{doi:10.1080/00140139208967394, doi:10.1080/00140130802170387, doi:10.3102/0034654317722961,DELGADO201823}.
Besides, notebooks are composed of cells (text or code) that can be read and executed in any order, which can lead to non-deterministic notebook behavior.
This non-determinism, which has been widely studied in the context of data science~\cite{10.1145/3313831.3376729, 10.1145/3290605.3300500, 10.1145/3173574.3173748, 10.1145/3173574.3173606}, can be very problematic in an educational context and must be carefully taken into account by the teacher instructions.

Notebooks offer a unique opportunity to provide more integrated instructions since all the material required for the learning activity is at the fingertip of the students on the same platform.
Building on this idea, we designed MOON, an approach intended to operationalize and enhance the efficacy of the instructions written by a teacher in a notebook-based learning activity.
This approach is enabled by two main contributions.
First, we provide a DFA-based language that allows teachers to express in a script how students are supposed to manipulate the different cells of a notebook.
Second, we materialize the script with a color system complemented by emojis that assists students directly inside notebooks (Figure~\ref{fig:student_notebook}).
The purpose of this assistance is to enable students to maintain a consistent notebook without preventing them from engaging in exploratory activities.

To evaluate the usefulness of our approach, we conducted a randomized controlled study involving first-year computer science students at the university in a three hours programming practical session on graph theory.
We observe that students in the control group (without MOON) complete as many cells as those using MOON. However, students in the control group have much greater difficulty following the written instructions provided by the teacher.
We also proceeded to a user study that confirms the findings of our controlled experiment and shows that most students found MOON useful.
The implementation of MOON, the notebooks used in the evaluation, and the raw data of the controlled study and user study are available in the accompanying Zenodo artifact~\cite{zenodo}.

%% file: description.tex
\section{Educational Notebooks}
\label{sec:notebooks}

\begin{figure*}[htb]
	\centering
	\begin{subfigure}{0.49\linewidth}
		\centering
		\includegraphics[width=\linewidth]{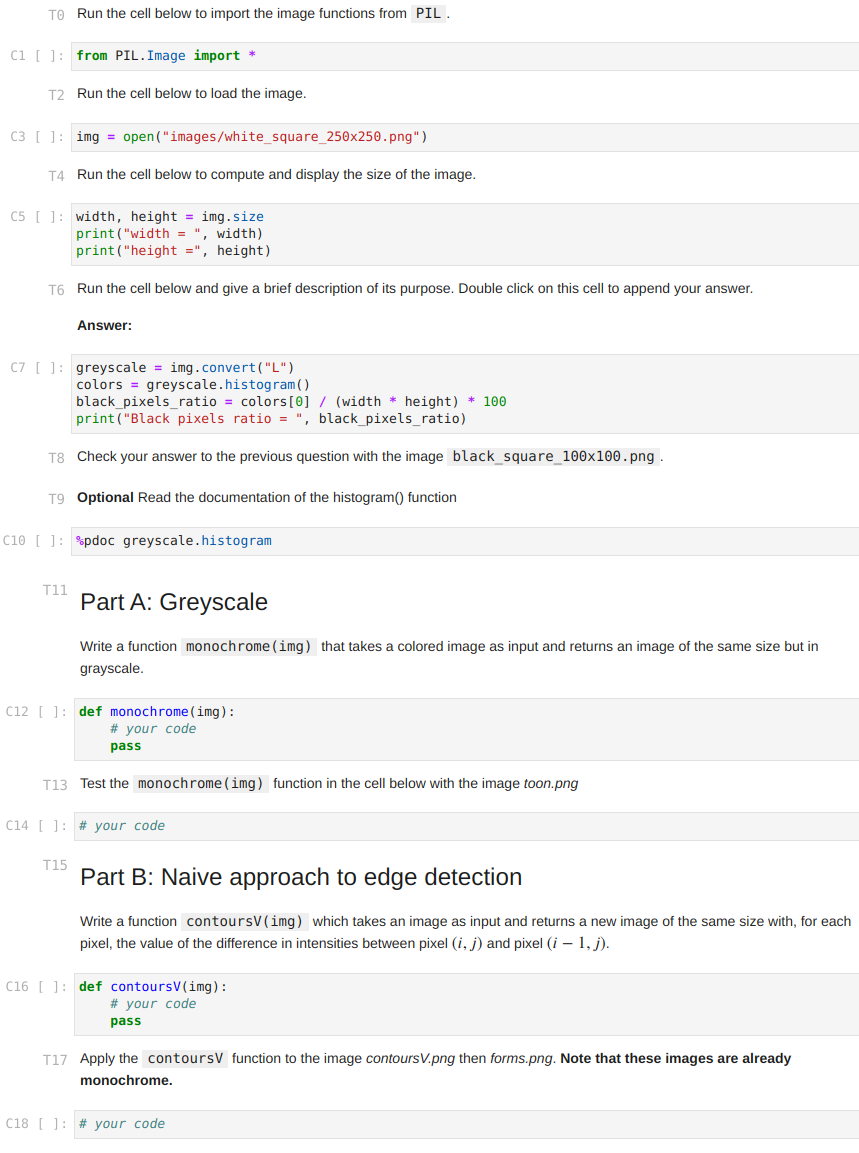}
		\caption{Example of an educational notebook embodying the teacher's scenario.}
		\label{fig:teacher_notebook}
	\end{subfigure}\hfill
	\begin{subfigure}{0.49\linewidth}
		\centering
		\includegraphics[width=\linewidth]{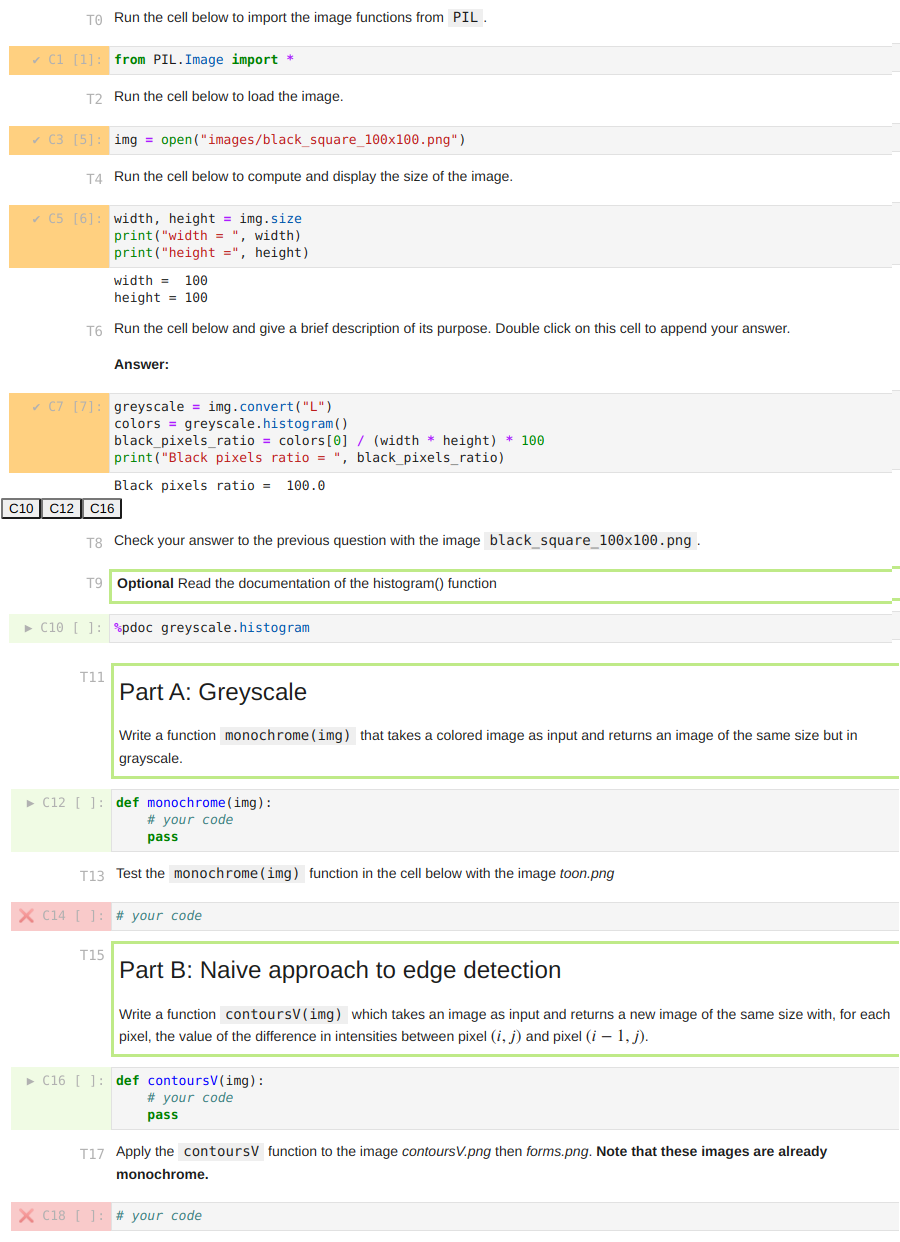}
		\caption{State of the notebook after the student is run in the following order the cells : \texttt{C1 C3 C5 C7 C3 C5 C7}}
		\label{fig:student_notebook}
	\end{subfigure}
	\caption{All cells of the notebook are labelled \texttt{Ci} for code cells and \texttt{Tj} for text cells, where \texttt{i} and \texttt{j} represent the indices of the notebook cells.}
	\label{fig:notebooks}
\end{figure*}
% Notebook scenario
There are multiple ways to use notebooks for teaching.
Our study primarily focuses on interactive educational notebooks used as course and exercise sheets, exams and assignments.
To create them, the teacher must write a notebook containing instruction cells related to the learning activity, as well as the instructions that will guide students in the notebook's execution model.
The written content constitutes the notebook's scenario.
Consider as an example an introductory image processing course in Python where students are asked to experiment with \texttt{Pillow}, an imaging library, and to implement well-known algorithms.
The teacher decides to organize the session around three parts.
Note that we present a simplified scenario compared to the one given to the students.
In the first part, students should import the library, load a sample image, compute its size and black pixel ratio, and try again with another image.
In the second part, students should implement a simple algorithm to convert an image to grayscale and showcase its use.
In the third part, students should implement a naive algorithm for edge detection and showcase its use.

% Scenario to notebook
Once the learning activity is finalized, the teacher materializes it in the notebook intended to be used by students during the session.
The challenges when writing the notebook's scenario are twofold: i) crafting text that effectively imparts knowledge and ii) providing clear directions for students to execute the code cells in the intended sequence.
Indeed, since notebook cells may be executed in any order, students may encounter issues if they attempt to use variables and functions before they are defined, for instance.
The implications for learning are significant, as students may confront execution errors and exceptions that divert their attention---and the teacher's!---from the pedagogical objectives.
Once the notebook is completed, the text and code cells should make it straightforward for students to follow the scenario written by the teacher.
%We call it an educational notebook.

% Illustration
\Cref{fig:teacher_notebook} depicts an educational notebook with a scenario.
For the sake of readability, each cell is associated with a unique identifier corresponding to its position in the notebook.
The letter \texttt{C} denotes code cells while the letter \texttt{T} denotes text cells.
As the instructions in the first part suggest, the students are asked to execute cells \texttt{C1}, \texttt{C3}, \texttt{C5}, and \texttt{C7} sequentially and then optionally execute cell \texttt{C10}.
After executing \texttt{C7}, the students should edit \texttt{C3} to load another image and re-run the intermediate cells \texttt{C5} to check their answers.
The remainder of the notebook is divided into two parts which can be worked on independently.

% Issues for students
When students open the educational notebook, they must read the text cells to follow the scenario.
Many studies have delved deeply into reading comprehension models and have shown that comprehension depends on many factors, such as reader characteristics, text content and design, and reading instructions~\cite{MCNAMARA2009297}.
A recent study even clearly demonstrates that reading comprehension scores are lower for digital texts than for print texts, regardless of age~\cite{DELGADO201823}.
Even with the best intentions, students may inadvertently end up with a notebook state not intended by the teacher.
Imagine a notebook in which the student executes cells \texttt{C1}, \texttt{C3}, \texttt{C5}, and \texttt{C7} sequentially and then changes the image used in cell \texttt{C3}, as requested.
However, they do not re-execute cell \texttt{C5}, thinking it is only intended to display the image's dimensions, and instead execute \texttt{C7} directly to compute the black pixel ratio.
The obtained result is incorrect because the black pixel ratio for the \emph{black\_square\_100x100.png} image is calculated with the height and width dimensions of the \emph{white\_square\_255x255} image that has not been updated (cell C5).
Without immediate feedback on the issue, there is a high risk that the student will not notice the issue and proceed, which may trigger other problems down the line and hurt their learning.
Finally, because notebooks foster live and exploratory programming, students may insert new code cells anywhere in the notebook to try things out, delete cells, or modify existing cells without ensuring the consistency of the entire notebook.
It is thus easy for students to corrupt their notebooks, sometimes without noticing, which hurts the consistency of the results and the intended scenario.

% Solution
To address these issues and support students in manipulating educational notebooks, this paper introduces a novel approach, MOON, fully integrated with Jupyter, that provides the teacher with a simple language for scripting their scenario within the notebook.
The resulting scripts are materialized while students interact with their notebooks, using visual assistance to guide them at every step by identifying the following cells to execute.

%% file: system_design.tex
\section{MOON Design}

Our approach, MOON, has two main objectives: i) to allow the teacher to express the scenario they have designed for their notebook in the form of a script and ii) to interpret the script to enrich the notebook with visual indications for the student identifying the next cells to be executed and information on past or upcoming executions of all the cells.

\subsection{From Scenarios to Scripts}

A scenario is embodied by the code and text cells of the educational notebook.
To better understand the execution patterns of code cells used by teachers in practice, we conducted an informal experiment.
We analyzed about one hundred Jupyter notebooks on GitHub and Kaggle that included instructions for the reader.
Our corpus of notebooks primarily consisted of educational notebooks, mostly from university courses.
These notebooks were selected to cover a wide range of academic disciplines and subjects, to try and have comprehensive collection of educational materials (exercices worksheets, assignments and interactive textbooks).
We identified three main patterns for cell execution: i) linear execution, ii) non-linear execution, and iii) optional execution.
We then created a simple scripting language that covers these three execution models and allows the teacher to define the intended sequence of execution of code cells in their notebook.
We limit ourselves to these three patterns as they are sufficient to express all the scenarios we encountered.

The basic construct manipulated in our scripting language is the code cell, each optionally associated with a set of text cells containing the corresponding instructions.
Code cells are written as follows: \texttt{Ci$\sim$Tj$\sim$...$\sim$Tn}, where $i, j, n \in \mathbb{N}$ denote the indices of the cells involved in the notebook.
For example, in the notebook of \Cref{fig:teacher_notebook}, \texttt{C1$\sim$T0} denotes the code cell \texttt{C1} and its associated instructions in text cell \texttt{T0}.
In the rest of the article, we omit textual cells from the scripts for the sake of clarity.
The execution order for the entire notebook is then specified by combining cells using three operators derived from the execution patterns above:

\paragraph{The linear pattern}
The linear pattern is naturally the most common, as it matches the standard top-down reading order.
To express that a set of code cells should be executed linearly, our script language uses the parentheses operator \texttt{()}.
The first part of the teacher's scenario involves a linear execution pattern that requires the following cells to be executed sequentially: \texttt{C1 C3 C5 C7 C3 C5 C7}.
Using the scripting language, this is expressed as \texttt{(C1 C3 C5 C7 C3 C5 C7)}.
\paragraph{The non-linear pattern}
The non-linear pattern indicates that a set of cells may be executed in any order, using the square brackets operator \texttt{[]}.
For example, with two code cells noted \texttt{Ci} and \texttt{Cj} we write \texttt{[Ci Cj]} which gives two possibilities of execution \texttt{Ci} then \texttt{Cj} or \texttt{Cj} then \texttt{Ci}.
\paragraph{The optional pattern}
The optional pattern, denoted with a question mark, indicates that a (set of) code cells may or may not be executed.
In our example, the first part of the notebook is expressed as \texttt{C1 C3 C5 C7 ?C10}, with \texttt{C10} optional.

Most importantly, the scripting language makes it possible to compose these patterns.
To illustrate this combination of operators, consider the notebook of \Cref{fig:teacher_notebook} in its totality.
The student must start with the first part, and then the remaining two (Part~A and Part~B) may be done in any order.
The first part is written as \texttt{(C1~C3~C5~C7~C3~C5~C7~?C10)} in the script with \texttt{C10} an optional cell. 
The script for parts A and B are as follows: A$\rightarrow$\texttt{(C12~C14)} and B$\rightarrow$\texttt{(C16~C18)}.
The three parts are put together by combining their scripts with the linear and any order patterns, as follows:~\texttt{((C1~C3~C5~C7~C3~C5~C7~?C10)[(C12~C14)(C16~C18)])}.
Parenthesis or bracket operators can be prefixed with a question mark to indicate optional cells.
%It is also possible to prefix the parenthesis or bracket operators with a question mark to mark all the cells they encompass as optional.

\subsection{Assisting Students in their Manipulation of the Notebook}
\label{sec:colors}

Once the teacher has written the script implementing their scenario, it is fed into MOON to guide students while they are working on their notebooks.
MOON uses a three-color coding system to guide students in realizing the scenario.
Specifically, at every step, MOON highlights the cells that can be executed in green, cells that have been completed in orange and cells that are not yet ready for execution in red.
In the example depicted in~\Cref{fig:student_notebook}, we can see that cells \texttt{C1}, \texttt{C3}, \texttt{C5} and \texttt{C7} are colored in orange, indicating that the student has already executed them.
MOON indicates that three cells, highlighted in green, can now be executed:~the optional cell \texttt{C10$\sim$T9};~Part~A with the code cell \texttt{C12$\sim$T11}; and Part~B with the code cell \texttt{C16$\sim$T15}.
Note that associated text cells are also highlighted to guide the student to the cells that contain the instructions for the different tasks that are immediately accessible.
If the student ignores the optional code cell \texttt{C10}, it turns red at the next step, whereas if they execute it, it turns orange.
Tasks that are not yet accessible and should not be executed are colored in red.
If a student executes a red cell, the colors of the cells remain unchanged.
The only way to progress in the scenario and highlight the next set of green cells is to execute one of the green cells that denote a task.
These visual indicators inform the students about their progression in the notebook.

Additionally, MOON decorates the last executed cell with buttons pointing to the next possible cells to execute.
This feature serves two primary objectives: i) to give an overview of the next possible cells without having to scroll through the page if they are outside the visible screen area; ii) to allow the students to quickly navigate through the notebook by clicking on the button corresponding to the desired code cell.
In our example, there are three buttons to indicate and jump to the three cells that can now be executed, corresponding to the optional cell (\texttt{C10}), Part~A (\texttt{C12}), and Part~B (\texttt{C16}).

\subsection{Support for Exploratory Programming}
\label{sec:exploratory}

Jupyter notebooks are a powerful tool for education due to their support for exploratory programming, enabling students to learn and experiment with code with great flexibility.
At first glance, MOON considerably reduces this flexibility by forcing students to follow the teacher's scenario.
However, to still encourage exploration, MOON accounts for the cases where students re-execute past cells and insert new cells while keeping the intended scenario on track.

\subsubsection{Automatic Backtracking}

In the exploratory phase, students may modify and re-execute the cells they have already executed, typically to correct their code or explore alternatives.
In MOON, this means that a student may execute an orange cell which is not one of the cells planned in the scenario at this point.
These exploratory phases are necessary for the students but potentially dangerous for the execution model of the notebook, as shown in the example of \Cref{fig:student_notebook}.
In this case, MOON offers an \emph{automatic backtracking} menu that assists the student in resuming the teacher's scenario as seamlessly as possible.
When executing an orange cell, MOON puts the students back to the last execution of that cell intended by the script and updates the colors of all cells in the notebook accordingly.
While the colors assigned to the different cells will be updated, the memory state of the notebook will not, and the students must manage the potential issues.

Looking back at our example with the following script \texttt{(C1~C3~C5~C7~?C10)}, when the student decides to use the image of the black square to check the ratio of black pixels, they forget to validate cell \texttt{C5}, yielding surprising results for cell \texttt{C7}.
With MOON, the execution of the orange code cell \texttt{C3} puts the student back in the script by coloring cell \texttt{C5} in green again.
The execution of cell \texttt{C5} allows obtaining the update of the variables \texttt{width} and \texttt{height} of the image \emph{black\_square.png} and the green highlighting of code cell \texttt{C7}.
When the student executes cell \texttt{C7}, they now obtain the correct result for the ratio of black pixels.
Finally, MOON also includes a \emph{back} button that allows students to go back one step in the sequence of the script.

\subsubsection{Adding and Deleting Cells}

MOON retains the ability to add new code and text cells to support active reading.
For example, students can create additional code cells beyond those provided by the teacher, which can serve as exploratory cells, or create text cells that can be used for note-taking, allowing them to keep all their work in a single document.
When one of these cells is created, it remains white to distinguish it from the cells in the scenario.
MOON also adjusts the indices of cells belonging to the scenario in the script, if necessary.
However, none of the cells involved in the scenario can be deleted once the script has been loaded.

%% file: implementation.tex
\section{Implementation}
\label{sec:implementation}

\begin{figure}[tb]
\begin{tikzpicture}[>={Stealth[round]},shorten >=1pt,node distance=1.8cm,on grid,auto,every state/.style={minimum size=0pt},bend angle=75]

  \node[state]          (q_0)                      {$q_0$};
  \node[state]          (q_1) [right=of q_0]       {$q_1$};
  \node[state]          (q_2) [right=of q_1]       {$q_2$};
  \node[state]          (q_3) [above right=of q_1] {$q_3$};
  \node[state]          (q_4) [below right=of q_1] {$q_6$};
  \node[state]          (q_5) [right=of q_3]       {$q_4$};
  \node[state]          (q_6) [right=of q_5]       {$q_5$};
  \node[state]          (q_7) [right=of q_4]       {$q_7$};
  \node[state]          (q_8) [right=of q_7]       {$q_8$};
  \node[state]          (q_9) [above right=of q_8]   {$q_9$};

  \path[->,every node/.style={font=\scriptsize}] (q_0) edge              node        {\texttt{C7}} (q_1)
  
            (q_1) edge              node                       {\texttt{C10}} (q_2)
                  edge [loop below] node                       {\texttt{C7}} ()
            (q_1) edge              node                       {\texttt{C12}} (q_3)
            (q_1) edge              node [pos=0.3, right=0.05] {\texttt{C16}} (q_4)
            (q_2) edge              node [right]               {\texttt{C12}} (q_3)
                  edge [loop right] node                       {\texttt{C10}} ()
            (q_2) edge              node [pos=0.2]             {\texttt{C16}} (q_4)
            (q_3) edge              node                       {\texttt{C14}} (q_5)
                  edge [loop above] node                       {\texttt{C12}} ()
            (q_4) edge              node                       {\texttt{C18}} (q_7)
                  edge [loop below] node                       {\texttt{C16}} ()
            (q_5)   edge              node                     {\texttt{C16}} (q_6)
                  edge [loop above] node                     {\texttt{C14}} ()
            (q_6)   edge              node                     {\texttt{C18}} (q_9)
                    edge [loop above] node                     {\texttt{C16}} ()
            (q_7)   edge              node []                   {\texttt{C12}} (q_8)
                    edge [loop below] node                      {\texttt{C18}} ()
            (q_8)   edge              node[pos=0.2, right=0.05] {\texttt{C14}} (q_9)
                    edge [loop below] node                     {\texttt{C12}} ();
\end{tikzpicture}
\caption{Automaton obtained from the script: \texttt{(C7 ?C10 [(C12~C14)(C16 C18)])}}
\label{fig:automaton}
\end{figure}
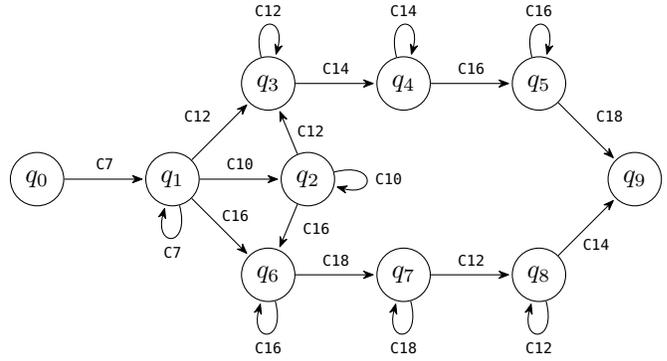

% DFA compilation
MOON is implemented as a JupyterLab plugin that takes as input a script as defined in the previous section. 
Note that our scripts are regular expressions, except for non-linear patterns.
We convert these patterns to obtain all the possible combinations of corresponding linear patterns and produces a deterministic finite automaton (DFA) using the algorithm described in ~\cite{10.5555/1196416}.
Therefore, it is not possible to have a scenario consisting of too many cells that could be executed in any order due to exponential complexity.
To construct the AST of a script, we use tsPEG\footnote{https://github.com/EoinDavey/tsPEG}, a PEG Parser generator designed for TypeScript. 
In the context of a notebook, the DFA is a model that represents a set of states and transitions between them, based on code cell executions.
If we take our notebook as an example~(Figure~\ref{fig:teacher_notebook}) and the following script: \texttt{C7 ?C10 [(C12 C14) (C16 C18)]}, we build the DFA shown in Figure~\ref{fig:automaton}.

% User trace
In addition to the DFA, we also maintain a user trace which is a sequence of pairs $(\text{\texttt{Ci}}, q_j)$ that corresponds to the list of valid transitions \texttt{Ci} leading to a state $q_j$ done by a given user.
Note that invalid transitions are not recorded in this trace.
For instance, if a user executes \texttt{C7}, \texttt{C12} and \texttt{C18} for the previously described notebook and script and starting at $q_0$, the user trace will contain: $(\text{\texttt{C7}}, q_1), (\text{\texttt{C12}}, q_3)$.

% Coloring process
When the user executes a code cell, the automaton checks if this execution corresponds to a possible transition for the current state.
If the transition is allowed, the automaton changes state and the colors of the notebook cells (code and text) are updated.
For our example~(Figure~\ref{fig:automaton}), consider the automaton in state $q1$ with a user trace containing $\text{\texttt{(C7}},q_1)$.
All cells belonging to the user trace are orange.
There are three transitions allowed by the automaton.
The execution of the code cell \texttt{C10} which is optional and the execution of the code cells \texttt{C12} and \texttt{C16} in any order.
These three code cells, which represent the next possible tasks, are green, as are their associated text cells and the other code cells are red (Figure~\ref{fig:student_notebook}).
Let's now look at what happens depending on whether or not the optional code cell \texttt{C10} is executed.
If the user validates the transition associated with the cell \texttt{C10}, it is added to the trace $\text{\texttt{(C7}},q_1),\text{\texttt{(C10}}, q_2)$ and turns orange.
In this case, there are now two possible transitions: the execution of code cells \texttt{C12} and \texttt{C16}, whose cells are colored green on the notebook. The other code cells are red.
If the user validates the transition associated with the code cell \texttt{C12} without having validated the transition associated with the code cell \texttt{C10}, the user trace becomes $\text{\texttt{(C7}},q_1),\text{\texttt{(C12}}, q_3)$ and the code cell \texttt{C12} becomes orange.
In this case, the next authorized transition is associated with code cell \texttt{C14} (green) and the code cells \texttt{C10} and \texttt{C16} turn red on the notebook. 

% Backtracking process
We also implemented the possibility for the student to re-run an orange cell and place the automaton in the state that corresponds to this transition in the script.
For example, if we have the following user trace: $\text{\texttt{(C7}},q_1),\text{\texttt{(C12}}, q_3),\text{\texttt{(C14}}, q_4)$, on the notebook the next task is designated by the green code cell \texttt{C16}.
Imagine now, that the student changes the orange code cell \texttt{C12} and then executes it.
The automatic backtracking looks in the user trace for the last pair containing the transition associated with the \texttt{C12} code cell by deleting all the next ones in the trace.
Finally the user trace contains only $\text{\texttt{(C7}},q_1),\text{\texttt{(C12}}, q_3)$ and the DFA is resetted to state $q_3$.
The code cells \texttt{C7} and \texttt{C12} are orange, the code cell \texttt{C14} is green and other code cells are red.

Let's illustrate another possible situation with the following new user trace: $\text{\texttt{(C0}},q_1),\text{\texttt{(C2}}, q_2),\text{\texttt{(C0}}, q_3),\text{\texttt{(C4}}, q_4)$, with a DFA in state $q_4$.
From this trace, we can observe that the scenario requires the student to execute code cell \texttt{C0} twice, first at the beginning and then after executing code cell \texttt{C2}.
If the student decides to execute code cell \texttt{C0} again, the backtracking implementation will reset the DFA to state $q_3$ and code cell \texttt{C4} will turn green.
However, with backtracking alone, it is not possible to reset the DFA to state $q_1$.
To overcome this limitation, we have also implemented a "back" button that allows users to reset the DFA to the previous state in the user trace.
Each click on the back button removes the last pair in the user trace, resets the DFA to the previous state, and updates the cell colors in the notebook.
Once the DFA is in state $q_4$, it is sufficient to click the back button four times to reset the DFA to state $q_1$.
In a more complex user trace, it is also possible to combine automatic backtracking with the back button to return to a previous state.

\noindent

%% file: experimental_design.tex
\section{Controlled Experiment}
\label{sec:study-design}

% Research questions
To evaluate the ability of MOON to support students in manipulating educational notebooks with a scenario, we conducted a controlled experiment.
Specifically, our evaluation aims to answer the two following research questions:
\begin{description}
	\item[RQ1] Does MOON help the students to better comply with an educational notebook with a scenario?
	\item[RQ2] Does MOON hinder the ability of students to progress in the scenario?
\end{description}

Our hypothesis for RQ1 is that the assistance provided by MOON might decrease the amount of incorrect usage of the notebook.
Our hypothesis for RQ2 is that the assistance provided by MOON might make it longer to manipulate the notebook and hinder the students' progression.

\subsection{Experimental Design}

% Description of the experiment
To study the impact of MOON, we proceeded to an A/B testing controlled experiment, where one group of students is given a notebook without MOON, and the other group is assigned the same notebook with MOON support.

% Course context
We conducted our experiment as part of an introductory programming course involving 21 first-year bachelor students from the university of Bordeaux during the first semester.
This course is organized around a set of chapters (\eg arrays, sorting algorithms, images, graph theory). Each chapter is implemented as a Jupyter notebook file which interweaves course concepts, images, source code, and coding questions.
The study took place in the first week of January 2023, the students had already manipulated notebooks during 6 x 2H40 practical sessions, and the study took place in the 7th (last) session. Hence, the students already had experience with notebooks.

% Group constitution
For the experiment, we randomly assigned 11 students to the control group and 10 students to the experimental group that uses MOON.
The students using MOON received a short twenty-minute training session where one of the authors, who was the instructor of the students, explained the meaning of the colors (\Cref{sec:colors}) and the features of the backtracking menu (\Cref{sec:exploratory}) on a sample notebook.

% Experiment notebook 
The notebook used in this experiment consists of 3 exercises on graph theory with different cell execution patterns~\cite{zenodo}.
The first exercise follows a linear execution pattern with an optional cell.
The second exercise combines linear and non-linear execution patterns.
The third exercise only involves a linear execution pattern.
The last two exercises can be done independently, so the student may start with either one.
The complete notebook requires 23 cell executions distributed across 20 code cells and 23 markdown cells:~three code cells are expected to be executed at least twice.
Each code cell requires the student to either execute, complete, or fill in the cell.
This notebook was distributed to the students during a 2h40 practical session.
The session's instructor was tasked to only answer questions related to the course's content (graph theory) and not to assist students in manipulating the notebook to avoid biasing their behavior.
At the end of the practical session, the instructor gathered the notebooks produced by the students.

% Notebook monitoring
To answer our research questions, we monitored how the students manipulate the notebook.
To that extent, we instrumented JupyterLab for both groups to record a complete trace of the students' cell execution actions on the notebook.
This trace, the log trace, is saved into the metadata of the notebooks.
Note that the log trace differs from the user trace described in \Cref{sec:implementation} because i)~it records every cell execution, even those not complying with the scenario, and ii)~it is never modified, even in case of backtracking.
Finally, we simplify the log traces by removing and replacing sequences of execution of the same cell with a single execution of this cell.
We perform this simplification because it is common for students to execute multiple times the same code cell when working on it. Still, it does not affect compliance with the intended scenario.

% Metrics RQ1
To study RQ1, we introduce a metric to capture the number of incorrect usage of the notebook.
Our metric is therefore defined based on the concept of \emph{deviation} from the scenario, where a deviation is the execution of a red cell, \ie the execution of a cell that is not permitted in the scenario.
To compute whether the actions recorded in the log traces correspond to a valid (green) transition, a backtracking (orange), or an invalid transition (red), we replay the students' log traces offline with MOON after the practical session.
Finally, to define our metric, called \emph{fitness}, we note $g$, $o$, $r$ the total number of green, orange, and red cells executed in a student trace.
Fitness is defined as the ratio of green or orange cell executions (\ie valid executions) over the total number of cell executions: $\mathrm{fitness} = (g + o)/(g + o + r)$

A fitness of $1$ indicates that a student has never executed a red cell, while a value of $0$ indicates that a student only executed red cells.
To assess the differences between the control and experimental groups, we use a two-sided non-parametric Mann-Whitney U test with a $p$-value cutoff of $0.01$ and report the Rank Biserial Correlation as an effect size.
Our null hypothesis is that the fitness is the same in both groups, and our alternate hypothesis is that the fitness is different in both groups.

% Metrics RQ2
To study RQ2, we introduce a metric to capture how far the students went in the notebook.
To that extent, we define \emph{completeness} as the number of distinct cells executed in a given log trace.
A student that did not execute any cell of the notebook corresponds to a completeness of $0$, while a student that executed every possible cell inside the notebook has a completeness of $20$.
To assess the difference between the control and experimental groups, we use a two-sided non-parametric Mann-Whitney U test with a $p$-value cutoff of $0.01$, and we report the Rank Biserial Correlation as an effect size.
We also evaluate the students' code by assigning a grade to each notebook.
Our null hypothesis is that the completeness and/or correctness is the same in both groups, and our alternate hypothesis is that the completeness and/or correctness is different in both groups.

\subsection{Results}
\label{sec:results}

\subsubsection{RQ1}

\begin{figure*}
	\centering
        \begin{subfigure}{0.32\linewidth}
	  \includegraphics[width=\linewidth]{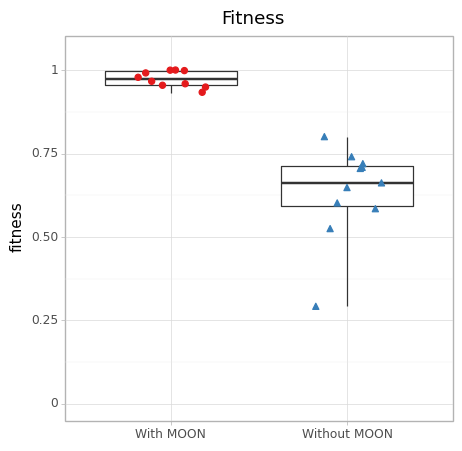}
	  \centering
	  \caption{Fitness in the control and experimental groups}
	  \label{fig:fitness}
        \end{subfigure}
        \begin{subfigure}{0.32\linewidth}
	\includegraphics[width=\linewidth]{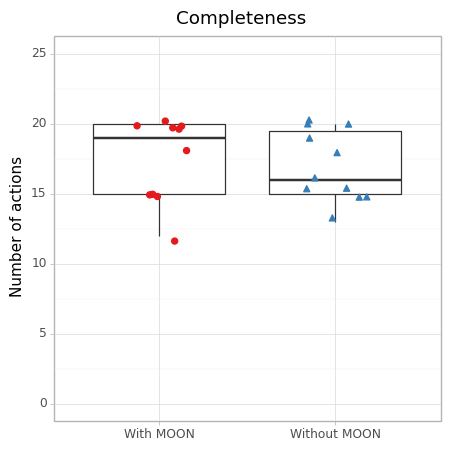}
	\centering
	\caption{Completeness in the control and experimental groups}
	\label{fig:completeness}
        \end{subfigure}
        \begin{subfigure}{0.32\linewidth}
	\includegraphics[width=\linewidth]{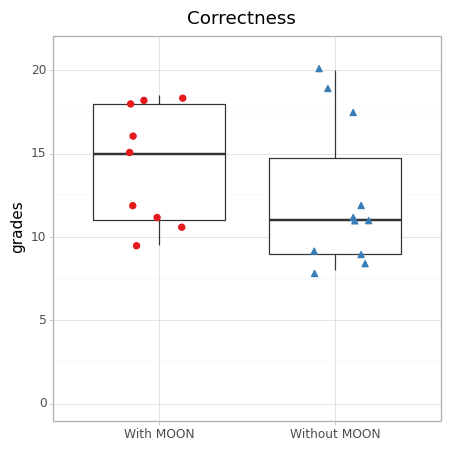}
	\centering
	\caption{Correctness in the control and experimental groups (french grading system [0-20])}
	\label{fig:grades}
        \end{subfigure}
        
        \caption{Analysis of student notebooks for experimental (with MOON) and control groups (without MOON).}
	\label{fig:user-study}
\end{figure*}

Figure~\ref{fig:fitness} depicts the fitness of the log traces from both groups.
Visually, we note a sharp difference between the two groups.
The fitness in the control group varies between about $0.25$ (meaning only a quarter of the executions are in a green or orange cell) to about $0.75$.
In the experimental group, the fitness values are very close to $1$ for all traces, with three traces having a fitness of $1$ (no execution of a red cell).
The Mann-Whitney U test yields a $p$-value $p=0.000122$ and an effect size $\mathit{rbc} = -1$.
We thus reject the null hypothesis and accept the alternate hypothesis.
The effect size indicates a clear effect in favor of the group using MOON, with no fitness value from the control group (maximum value $0.8$) exceeding the lowest fitness value in the experimental group ($0.91$).
In conclusion, the data support the hypothesis that MOON reduces the amount of incorrect notebook usage.

\subsubsection{RQ2}

Figure~\ref{fig:completeness} depicts the completeness computed from the log traces from both groups and figure~\ref{fig:grades} shows the students' scores for the practical session.
For this two figures, visually, the distribution of the values is similar between the two groups, with a comparable spread and a slightly lower median in the control group.
For completeness, the Mann-Whitney U test yields a $p$-value $p=0.53$ and an effect size $\mathit{rbc} = -0.145455$.
Thus, we cannot reject the null hypothesis that the completeness is the same between the groups.
For Correctness, the Mann-Whitney U test yields a $p$-value $p=0.29$ and an effect size $\mathit{rbc} = -0.292929$.
Thus, we cannot reject the null hypothesis that the grades are the same between the groups.
The calculation of the median reveals a 4-point difference in favor of the group with MOON.
Similarly, when calculating the average, the experimental group has a mean of 14.3 with a standard deviation of 3.6, while the control group has a mean of 12.4 with a standard deviation of 4.4.
Moreover, in both cases the effect size is small and in favor of the experimental group.
Therefore, the data does not support the hypothesis that MOON hinders students' progression.
On the contrary, there is a slight advantage in the experimental group.

In summary, based on our controlled experiment, we answer our research questions as follows:
\begin{tcolorbox}
MOON helps students to better comply with the notebook's scenario without hindering their ability to progress in the scenario.
\end{tcolorbox}

\subsection{Threats to Validity}

Regarding construct validity, the completeness metric might not accurately capture the actual progress of students inside the notebook, as executing a cell is different from completing the work required in this cell.
To counter this threat, the instructor reviewed all notebooks and ensured there were no discrepancies between the log trace and the completed cells in the 21 notebooks.
We examined the correctness of the students' notebooks, but we had reservations about the subjectivity of grading since it was done by one of the authors.

Regarding internal validity, JupyterLab's ability to restart the execution kernel can severely disrupt the use of MOON since it resets the notebook's memory while keeping the state of MOON.
Therefore, the instructor asked the students to reset MOON if they had to restart the kernel, but some students could have ignored this instruction.
The instructor also advised students to only click on the refresh button on the browser if they were sure they had saved their notebooks.
This action resets the plugin and clears the user trace to start a new one.
MOON makes heavy use of colors to communicate feedback to students, which might be an issue for some students due to \eg color blindness.
To combat the threat, emojis were added with the colors.
In the experimental group, no students had issues using MOON.

Regarding external validity, we performed our controlled experiment in only one context. Therefore, the results could vary with level students or the notebook used in the session.

%% file: user_study.tex
\section{User Study}
\label{sec:user-study}

\begin{figure*}
	\centering
	\begin{subfigure}{0.24\linewidth}
		\centering
		\includegraphics[width=\linewidth]{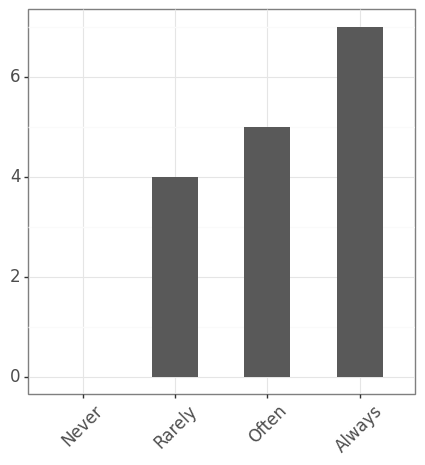}
		\par\bigskip
		\caption{Does MOON help to follow the execution order of cells?}
		\label{fig:moonorder}
	\end{subfigure}
	\begin{subfigure}{0.24\linewidth}
		\centering
		\includegraphics[width=\linewidth]{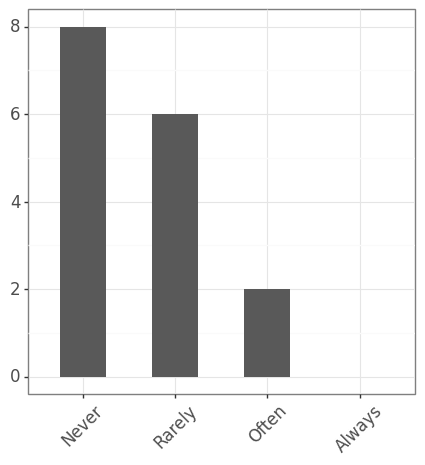}
		\par\bigskip
		\caption{Are the actions suggested by MOON surprising?}
		\label{fig:moonactions}
	\end{subfigure}
	\begin{subfigure}{0.24\linewidth}
		\centering
		\includegraphics[width=\linewidth]{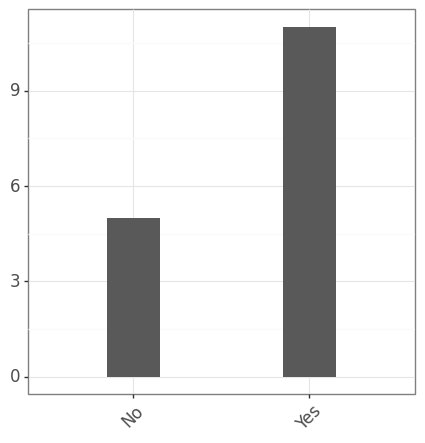}
		\par\bigskip\medskip
		\caption{Does MOON help to progress faster?}
		\label{fig:moonspeed}
	\end{subfigure}
	\begin{subfigure}{0.24\linewidth}
		\centering
		\includegraphics[width=\linewidth]{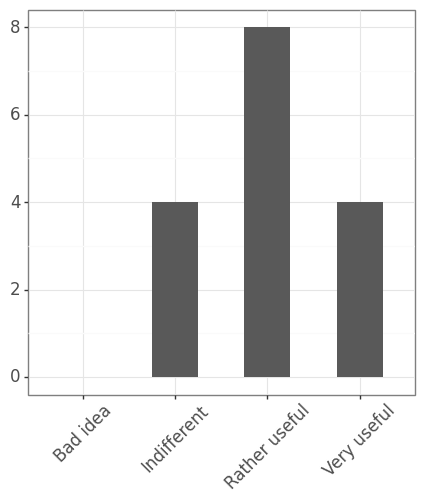}
		\caption{Do you think that MOON should be activated by default?}
		\label{fig:moonopinion}
	\end{subfigure}
	\caption{Distributions of the answers for the questions of our anonymous survey}
	\label{fig:user-study}
\end{figure*}

To complement our controlled experiment, we asked the 10 students involved in the experimental group and 10 other students who were involved in the beta testing of MOON to complete an anonymous online survey about MOON.
We received 16 answers to this survey yielding a response rate of $80\%$.
\Cref{fig:user-study} depicts the distribution of the answers to the four main questions.

To the question ``\emph{Does MOON help to follow the execution order of cells?}'' we observe the distribution depicted in Figure~\ref{fig:moonorder}.
We note that the most frequent answer is \emph{Always}, with 7 answers.
We also note that no participant answered \emph{Never}.
This result corroborates what we observed for RQ1 in the controlled experiment, where the fitness observed in the experimental group was significantly higher than in the control group.

To gain a deeper understanding of how the students perceive the actions suggested by MOON, we asked the question ``\emph{Are the actions suggested by MOON surprising?}''
Figure~\ref{fig:moonactions} depicts the answers to this question.
We note that the most frequent answer is \emph{Never}, with 8 answers, and only two participants answered \emph{Often}.
It indicates that students mostly understand the actions suggested by MOON, but some actions might be surprising in some cases.
We included a free-text question to add more context to this question.
Two students mentioned that they were surprised to have to validate the intermediate cells in case of backtracking, even if these cells were not modified.
One student mentioned that they were surprised to be able to execute a cell very distant from the current one.

To corroborate the results of RQ2 in the controlled experiment, we asked ``\emph{Does MOON help to progress faster?}''
Figure~\ref{fig:moonspeed} depicts the answers to this question.
We note that the most frequent answer is \emph{Yes} with 11 answers, and only 5 students answered \emph{No}.
This aligns with the fact that we did not see any effect of MOON on completeness, and we observed increased median completeness in the experimental group.

To assess whether the students found MOON helpful, we asked the question ``\emph{Do you think that MOON should be activated by default?}'', \ie, would the students like to use MOON in all their practical sessions?
Figure~\ref{fig:moonopinion} depicts the answers to this question.
Inspired by related work~\cite{10.1145/2568225.2568233}, we decided to use an asymmetric survey response scale inspired by the Kano model suitable for evaluating preferences.
We note that the most frequent answer is \emph{Rather useful} with 8 answers, while 4 students answered \emph{Indifferent} and 4 \emph{Very useful}.
None of the students answered \emph{Bad idea}.
It indicates that most students find MOON useful, while some do not seem to see much benefit in using it.
We conjecture this might be true for the most advanced notebook users.

Finally, we included a final free-text question asking for any feedback about MOON.
One student mentioned that validating intermediate cells in case of backtracking is tedious.
Another mentioned that the help provided by MOON did not stand out at first, but when he returned to the notebook without it, the difference was apparent, and the notebook seemed more complicated to use.

%% file: related_work.tex
\section{Related work}

We discuss the related work around two topics: the use of notebooks in education and digital storytelling.

The literature shows that notebooks are increasingly used in teaching\footnote{https://www.epfl.ch/education/educational-initiatives/jupyter-notebooks-for-education/} and that there are mainly four ways of using them: video projection (including live programming~\cite{10.1145/3330430.3333627}), as course and exercise sheets~\cite{sutrini2022potential,doi:10.1021/acs.jchemed.0c01071, 9962690}, and to support automatic evaluation~\cite{nbgrader, 10.1145/3328778.3366947, su132112050}.
Our approach only encompasses the course and exercise sheets aspect which allows the student to have in the same document his code, the execution of the code and the instructions for an exercise session.
But the notebook written by the teacher is often put to the test when the students are in the exploratory phase~\cite{10.1145/3313831.3376729,johnson2020benefits}.
It is therefore necessary for the teacher to have a precise idea of how a notebook works, its advantages but also its pitfalls.
Once this knowledge has been acquired, the teacher must face a double challenge: writing instructions to guide the student in his or her learning process and acquiring good practices~\cite{wang2020better,quaranta2022eliciting} to avoid the execution pitfalls intrinsic to the notebook~\cite{10.1145/3173574.3173606, 10.1145/3408877.3432361}.

Several tools have been developed to improve the use of notebooks in the classroom, such as plugins for automatic evaluation~\cite{nbgrader}, reproducibility~\cite{casseau2021immediate,fangohr2020testing}, or the use of Personal Learning Environments such as GRAASP~\cite{inproceedings_graasp}, with the integration of a tool resembling the notebook~\cite{9790858}.

Several other approaches are not directly related to teaching but explore another very interesting aspect for the teacher related to storytelling by modifying the user interface of notebooks~\cite{li2023notable,kang2021toonnote,weinman2021fork,harden2022exploring,wang2022stickyland}.

Although these tools allow for improved interactions between the notebook and the users in terms of reproducibility, sharing, and storytelling, none of them are adapted for the teacher to embody his scenario in the notebook to assist with the reading and execution of it. We propose a new approach with MOON which allows the teacher to script his scenario in an educational notebook and to activate for the students' assistance the execution of the script based on a color system.

%% file: conclusion.tex
\section{Conclusion}
\label{sec:conclusion}

In this article, we present a novel approach, MOON, designed to guide students through educational notebook scenarios.
The central idea is to provide teachers with a language that enables them to formalize the expected usage of their notebooks in the form of a script and to interpret this
script to guide students with visual indications in real time while they interact with the notebooks.
We evaluate our approach using a randomized controlled experiment involving 21 students, which shows that MOON helps students comply better with the intended scenario without hindering their ability to progress.
Our follow-up user study shows that about 75\% of the surveyed students perceived MOON as rather useful or very useful.
In future work, we plan to investigate how we can leverage MOON to provide feedback to teachers.
A first idea would be to analyze the traces and the scripts to pinpoint pain points in a scenario and provide improvement suggestions.
A second idea would be to gather data from MOON instances in real-time to directly assist the teachers during a practical session by pinpointing students struggling with the notebook or lagging in the scenario.
Another important point to improve is script writing.
Currently, our approach imposes to write the script for the notebook scenario manually.
We are exploring ways to enhance the script writing process within the notebook scenario and ensure synchronization of the script when changes are made to the scenario.